\documentclass[11pt]{article}
\usepackage{epsfig}
\usepackage[usenames]{color}
\usepackage{amsmath}
\usepackage{amsfonts}
\usepackage{amssymb}
\usepackage{mathrsfs}
\newtheorem{proposition}{Proposition}[section]
\newcommand{\bpr}{\begin{proposition}}
\newcommand{\epr}{\end{proposition}}

\newcounter{Roman}
\addtocounter{Roman}{1}

\newcommand{\beq}{\begin{equation}}
\newcommand{\eeq}{\end{equation}}
\newcommand{\bea}{\begin{eqnarray}}
\newcommand{\eea}{\end{eqnarray}}
\newcommand{\bml}{\begin{multline}}
\newcommand{\bal}{\begin{align}}
\newcounter{saveeqn}

\newcommand{\ssc}{\scriptscriptstyle}

\newcommand{\bGamma}{\bar{\Gamma}}
\newcommand{\hGamma}{\hat{\Gamma}}

\newcommand{\bg}{\bar{g}}

\newcommand{\bnabla}{\bar{\nabla}}

\newcommand{\mf}[1]{\mathfrak{#1}}
\newcommand{\ms}[1]{\mathscr{#1}}

\input epsf
\begin{document}  

\begin{center}{\Large\bf 
On the `simple' form of the gravitational action and the self-interacting graviton
}
\\[2cm] 
{E. T. Tomboulis\footnote{\sf e-mail: tomboulis@physics.ucla.edu}
}\\
{\em  Mani L. Bhaumik Institute for Theoretical Physics\\
Department of Physics and Astronomy, UCLA, Los Angeles, 
CA 90095-1547}
\end{center}
\vspace{1cm}

\begin{center}{\Large\bf Abstract}\end{center} 
The so-called $\Gamma\Gamma$-form of the gravitational Lagrangian, long known to provide its most compact expression as well as the most efficient generation of the graviton vertices, is taken as the starting point for discussing General Relativity as a theory of the self-interacting graviton. 
A straightforward but general method of converting to a covariant formulation by the introduction of a reference metric is given. It is used to recast the Einstein field equation as the equation of motion of a spin-2 particle interacting with the canonical energy-momentum   
tensor symmetrized by the standard Belinfante method applicable to any field carrying nonzero spin. 
This represents the graviton field equation in a form complying with the precepts of standard field theory. It is then shown how representations based on 
other, at face value completely unrelated definitions of energy-momentum (pseudo)tensors are all related by the addition of appropriate superpotential terms. Specifically, the superpotentials are explicitly constructed which connect to: i) the common definition consisting simply of the nonlinear part of the Einstein tensor; ii) the Landau-Lifshitz definition. 

\vfill
\pagebreak

\section{Introduction} 
\setcounter{equation}{0}
\setcounter{Roman}{0}
As it is well-known since the early days of General Relativity (GR) \cite{Ed}, the Einstein-Hilbert (EH) action\footnote{In the following $\kappa^2= 16\pi G$, the Minkowski metric is denoted by $\eta= {\rm diag}(1, -1, -1, -1)$,  
and we use curvature conventions $R_{\mu\nu} = R^\lambda_{\mu\lambda\nu} = + \partial_\nu \Gamma^\lambda_{\mu\lambda} - \ldots$. When needed, 
functional dependence is, as usual, indicated by $[\,\cdot \, ]$, e.g., $S[\phi]$ for action functional $S$ of theory with fields $\phi$. \label{Fconv}}
\beq 
S_{\rm \ssc EH} 
= {1\over \kappa^2} \int d^4x \, \sqrt{-g} R(g)        \label{EHact}
\eeq 
can be written in  the form
\beq 
S_{\rm \ssc EH}=  {1\over \kappa^2} \int d^4x \, [ - \mf{L}  + \partial_\alpha \mf{B}^\alpha ]   \label{EHact1}
\eeq
with 
\beq 
\mf{L} = \sqrt{-g} g^{\mu\nu} \left( \Gamma_{\mu\beta}^\alpha \Gamma_{\alpha\nu}^\beta - \Gamma_{\mu\nu}^\alpha \Gamma_{\alpha\beta}^\beta \right)    \label{Ldens1}
\eeq
and 
\beq
\mf{B}^\alpha = \sqrt{-g} \left( g^{\alpha\mu} \Gamma_{\mu\nu}^\nu - g^{\mu\nu}\Gamma_{\mu\nu}^\alpha \right)   \, .\label{bound1}
\eeq
Dropping the total divergence gives the so-called $\Gamma\Gamma$-form of the Lagrangian, which contains only first derivatives of the metric. Its variation  
gives the Einstein field equations. 
It is noteworthy that $\sqrt{-g} R(g) = \sqrt{-g} (\partial \Gamma - \partial \Gamma) + \mf{L}$, but, after the isolation of the boundary term,  it is $-\mf{L}$ that gives this $\Gamma\Gamma$-form of the Lagrangian. $\mf{L}$ satisfies homogeneity properties and may be viewed equally well as a function of either $g_{\mu\nu}$, or $g^{\mu\nu}$, or $\mf{g}^{\mu\nu}=\sqrt{-g} g^{\mu\nu}$, or $\mf{g}_{\mu\nu}=g_{\mu\nu}/\sqrt{-g} $, and their respective derivatives. We return to these points in a more general context below. 

Expressing the metric connections  (Christoffel symbols) $\Gamma$ in (\ref{Ldens1}) in terms of $g_{\mu\nu}$  gives 
\bea
S  & = &  {1\over \kappa^2} \int d^4x \,  (-\mf{L})    \nonumber \\
& = & 
{1\over 4 \kappa^2} \int d^4x  \sqrt{-g} \, (\partial_\alpha g_{\mu\nu}) \left[g^{\alpha\beta} g^{\mu\kappa}g^{\nu\lambda} - g^{\alpha\beta}g^{\mu\nu}g^{\kappa\lambda} - 2g^{\alpha\lambda}g^{\beta\nu}g^{\mu\kappa} + 2g^{\mu\nu}g^{\alpha\kappa}g^{\beta\lambda}   \right] (\partial_\beta g_{\kappa\lambda} )  \nonumber \\
& \equiv &  {1\over  \kappa^2} \int d^4x \, (\partial_\alpha g_{\mu\nu})\, K^{\alpha\mu\nu\beta\kappa\lambda} (g) \,(\partial_\beta g_{\kappa\lambda} )  \, .   \label{EHact2}
\eea
Though irrelevant for writing the action (\ref{EHact2}), for many manipulations the symmetrized form of $K^{\alpha\mu\nu\beta\kappa\lambda}$:  
\bml
K^{\alpha\mu\nu\beta\kappa\lambda}(g) = {1\over 4} \sqrt{-g} \, \left[ g^{\alpha\beta} ( g^{\mu(\kappa} g^{\lambda)\nu} - g^{\mu\nu}g^{\kappa\lambda} )\right. \\
\left. - (g^{\alpha(\lambda} g^{\nu)\beta} g^{\mu\kappa} +   g^{\alpha(\kappa} g^{\mu)\beta} g^{\nu\lambda})        + (g^{\kappa\lambda}g^{\mu(\alpha}g^{\beta)\nu} + g^{\mu\nu}g^{\kappa(\beta}g^{\alpha)\lambda}) \right]     \qquad \qquad \qquad    \label{K1} 
\end{multline}
should be used. 
(\ref{EHact2}) gives the most compact form known of the gravitational action. A very similar, somewhat even simpler form is obtained by using $\mf{g}^{\mu\nu}$ as field variables. 
Expanding about flat background metric, $g_{\mu\nu} = \eta_{\mu\nu} + h_{\mu\nu}$, all graviton vertices are then obtained by the expansion of the inverse metric $g^{\mu\nu}$ dependence in $K(g)$ in powers of $h_{\mu\nu}$. The free quadratic part 
\beq 
S_2 [\eta,h]= {1\over  \kappa^2} \int d^4x \, (\partial_\alpha h_{\mu\nu})\, K^{\alpha\mu\nu\beta\kappa\lambda} (\eta) \,(\partial_\beta h_{\kappa\lambda} )    \label{FPauli1}
\eeq
is the Fierz-Pauli action for a massless spin-2 particle. All vertices ($n\geq 3)$  are then recursively generated by differentiation of this free kinetic part of the action:
\beq 
S_n[\eta,h] = {1\over (n-2)!} \big(\partial_s^{(n-2)} S_2[\eta + s h, h]\big)_{| s=0} \, ,  \label{vert1}
\eeq 
which, via $\partial_s g^{\mu\nu}(s) = \partial_s (\eta + sh)^{\mu\nu}= - g^{\mu\alpha}(s) g^{\nu\beta}(s) h_{\alpha\beta}$, amounts to simple insertion of powers of $h$ in the kinetic term with no change of derivative terms. 
This basically simple structure of the gravitational vertices has been known for a long time.  

If one adopts a first order formalism with $\mf{g}$ and $\Gamma$ taken as independent variables, one notes that (\ref{Ldens1}) contains only up to cubic interactions. This was, of course, the starting point of Deser's seminal work \cite{De1} - \cite{De2} conclusively demonstrating that the self-coupling of a massless spin-2 particle bootstraps to GR. Conversely, taking (\ref{EHact2}) as the starting point, its simple structure allows, by simple Gaussian integration,  the introduction of an auxiliary field, in terms of which one reverts to the first order formalism, the auxiliary field being, of course, essentially $\Gamma$.  This has recently been revisited in \cite{CR}.\footnote{It is pointed out in \cite{CR} that (\ref{EHact2}) may be somewhat further simplified by a gauge choice involving both $g_{\mu\nu}$ and it inverse $g^{\mu\nu}$. Though this can give some simplification of the graviton vertices it introduces an infinite number of interaction vertices in the Fadeev-Popov ghost Lagrangian, 
a disaster for any actual amplitude computations.}

Now, $\mf{L}$ given by (\ref{Ldens1}), is not a scalar density and, correspondingly, $\mf{B}$ is not a vector density. 
This is as it should be since there does not exist a scalar density which depends solely on the metric and its first derivatives (see, e.g., \cite{LR} for a proof). As a result the covariance of many results involving $\mf{L}$ is not always apparent. 
Similar, and related, issues come up in the representation of GR as the theory of the self-interacting graviton, which will be our main focus in the following. This amounts to writing the Einstein field equations as the equation of motion of a spin-2 particle coupled to an appropriate interaction current, which in this case is, of course, the energy-momentum pseudotensor of the gravity field. The definition of the latter is highly non-unique. The most common textbook definition consists of simply shifting the nonlinear part of the Einstein tensor to the r.h.s. of the field equations. Other manipulations of the field equations lead to different representations. A frequently used one is the Landau-Lifshitz (LL) pseudotensor. These and other ad hoc definitions make, in particular, no reference to the canonical energy-momentum tensor definition. The existence of all these disparate, and apparently unrelated definitions has been the source of some confusion in the literature over the years. Different definitions should give the same result for the total field energy-momentum vector in asymptotically flat spacetimes, but independence of the choice of coordinates in specific computations is not obvious due to the lack of manifest covariance. 
This has in fact motivated the formulation of more abstract global geometric 4-momentum definitions in asymptotically flat spacetimes \cite{TWi} - \cite{AshH}.

Our purpose in this note is to do two things. First, to give a covariant formulation so that quantities such as $\mf{L}$ and pseudotensors are converted to actual tensors and tensor densities, and coordinate independence of all results is manifest. This can only be done with the introduction of some extra structure. In section 2 below it is shown how this can be accomplished in a straightforward but general manner by the introduction of a symmetric but otherwise arbitrary reference connection. If this connection is taken to be a metric one, a reference metric is introduced. 
Introduction of a reference metric has indeed being used before in special schemes as in  \cite{PP}, and references therein,  
for the purpose of verifying agreement between the LL and the global geometric definitions for the total field 4-momentum. The simple procedure below is very general and applies to any metric gravity theory. The reference metric 
is naturally identified with the background metric if the usual case of propagation over a background is considered. 

Our second and main goal is to relate various ways that the Einstein field equations can be written as the equation of motion of the self-interacting graviton. This we do in section 3 using the covariant formulation of section 2.
Doing so means giving a precise relation between various definitions of the gravitational energy-momentum tensor.  
Of particular interest here is a formulation that adheres to the standard procedures of field theory. The canonical energy-momentum tensor associated with the Lagrangian density $\mf{L}$ is not symmetric. This is a feature of any field with nonzero intrinsic angular momentum and can be symmetrized for any spin by a standard procedure (Belinfante symmetrization) \cite{Bel}, \cite{Rsf}. 
The explicit construction involves the addition of an appropriate  superpotential term to the canonical definition.\footnote{Another well-known example of modification by addition of  a superpotential is the ``improved" energy-momentum tensor \cite{CCJ} ensuring vanishing trace in (classical) scale invariant theories.}  
The gravitational field equations are then transformed as the field equation of a spin-2 particle coupled to the Belinfante-symmetrized canonical energy momentum tensor. We then proceed to show how other  representations of the field equations are related to the canonical one. We present several new results and explicitly construct, in particular, the superpotential terms that connect the standard textbook tensor and the LL tensor to the canonical one. The fact that such different definitions must be related by addition of appropriate superpotential terms was stressed in \cite{De3} in discussing some of the confusion in the literature concerning these matters. Here we give the non-obvious expressions for the superpotentials connecting these different, at face value unrelated definitions.

We summarize in section 4 and briefly remark on the possibility of generalizing the theory of the self-interacting graviton beyond GR.

\section{Covariant formulation}
\setcounter{equation}{0}
\setcounter{Roman}{0}
The basic step is to expressed the 
covariant derivative operator $\nabla$ of the metric connection $\Gamma(g)$, for which $\nabla g=0$, in terms of the derivative operator 
$\bnabla$ of another arbitrary symmetric connection $\bGamma$. 
For any tensor $T^{\beta_1\cdots \beta_k}_{\qquad\alpha_1\cdots\alpha_l} $ one has 
\beq
\nabla_\mu T^{\beta_1\cdots \beta_k}_{\qquad\alpha_1\cdots\alpha_l} = \bnabla_\mu T^{\beta_1\cdots \beta_k}_{\qquad\alpha_1\cdots\alpha_l} + \sum_i\hGamma^{\beta_i}_{\mu\gamma} T^{\beta_1\cdots\gamma\cdots \beta_k}_{\qquad \quad\alpha_1\cdots\alpha_l} - \sum_j\hGamma^\gamma_{\mu\alpha_j} T^{\beta_1\cdots\beta_k}_{\qquad \alpha_1\cdots\gamma\cdots \alpha_l}    \label{barcovder1}
\eeq 
with 
\beq
\hGamma^\alpha_{\mu\nu} = {1\over 2} g^{\alpha\sigma} \left( \bnabla_\nu g_{\mu\sigma} + \bnabla_\mu g_{\nu\sigma} - \bnabla_\sigma g_{\mu\nu}  \right)   \label{barcovder2}
\eeq
and $\bnabla$ denoting the covariant derivative with connection $\bGamma$. (\ref{barcovder2}) is obtained from the condition $\nabla g_{\mu\nu}=0$ which is then easily verified to hold by applying (\ref{barcovder1}) to $g_{\mu\nu}$. 
Note that, as it is manifest from (\ref{barcovder2}), $\hGamma$ is a tensor. Indeed, it follows from (\ref{barcovder1}) that the relation $\Gamma(g)= \bGamma + \hGamma$ must hold, i.e., $\hGamma$ is the difference of two connections and hence a tensor.

It is now straigtforward to express the Riemann tensor in terms of the $\bGamma$ connection starting from the defining relation 
\beq
\left( \nabla_\alpha \nabla_\beta - \nabla_\beta \nabla_\alpha \right)  \upsilon_\gamma = R^\delta_{\ \gamma \alpha\beta} \,\upsilon_\delta   \, ,  \label{Riem0} 
\eeq
where $\upsilon$ is any covector. Using (\ref{barcovder1}) then  (\ref{Riem0}) yields
\beq 
R^\rho_{\ \mu\nu\sigma} = \bar{R}^\rho_{\ \mu\nu\sigma}  + \bnabla_\sigma \hGamma^\rho_{\ \mu\nu} - 
\bnabla_\nu\hGamma^\rho_{\ \mu\sigma} + \hGamma^\alpha_{\ \mu\nu} \hGamma^\rho_{\ \alpha\sigma} - 
\hGamma^\alpha_{\ \mu\sigma} \hGamma^\rho_{\ \alpha\nu}  \, .    \label{Riem1}
\eeq
Here $\bar{R}^\rho_{\ \mu\nu\sigma}\equiv R^\rho_{\ \mu\nu\sigma}(\bGamma)$.  

Insert now $R_{\mu\nu}= R^\rho_{\ \mu\rho\nu}$, as given by 
(\ref{Riem1}), in (\ref{EHact}) and  note that  
\beq 
\int \sqrt{-g} g^{\mu\nu} \left[  \bnabla_\nu \hGamma^\rho_{\mu\rho} - 
\bnabla_\rho\hGamma^\rho_{\mu\nu}  \right] = 
\int \bnabla_\alpha \mf{B}^\alpha + \int \left[ - \hGamma^\rho_{\mu\rho} \bnabla_\nu \mf{g}^{\mu\nu}  + \hGamma^\rho_{\mu\nu} \bnabla_\rho \mf{g}^{\mu\nu}  \right]    \label{actder1} 
\eeq
with 
\beq
\mf{B}^\alpha = \sqrt{-g} \left( g^{\alpha\mu} \hGamma_{\mu\nu}^\nu - g^{\mu\nu}\hGamma_{\mu\nu}^\alpha \right)   \, .\label{bound2}
\eeq 
The $\bnabla \mf{g}$ factors in the square bracket on the r.h.s. in (\ref{actder1}) may be expressed in terms of $\hGamma$'s by using the fact that $\nabla \mf{g}^{\mu\nu} = 0$ together with (\ref{barcovder1}) to obtain the useful identity 
\beq
\left[ - \hGamma^\rho_{\mu\rho} \bnabla_\nu \mf{g}^{\mu\nu}  + \hGamma^\rho_{\mu\nu} \bnabla_\rho \mf{g}^{\mu\nu}  \right]  = - 2 \mf{L}   \, , \label{i1}
\eeq
where now we define 
\beq  
\mf{L} = \sqrt{-g} g^{\mu\nu} \left( \hGamma_{\mu\beta}^\alpha \hGamma_{\alpha\nu}^\beta - \hGamma_{\mu\nu}^\alpha \hGamma_{\alpha\beta}^\beta \right)  \, .  \label{Ldens2}
\eeq
By  (\ref{Riem1}), (\ref{actder1}), (\ref{i1}) then, and use of the divergence identity 
$\bnabla_\alpha \mf{B}^\alpha = \partial_\alpha \mf{B}^\alpha$, valid for any vector density, one obtains 
 \beq 
S_{\rm \ssc EH}=  {1\over \kappa^2} \int d^4x \, [ \sqrt{-g} g^{\mu\nu} \bar{R}_{\mu\nu}   - \mf{L}  + \partial_\alpha \mf{B}^\alpha ]  \, .  \label{EHact3}
\eeq
Here, $\mf{L}$ and $\mf{B}$, now given by (\ref{Ldens2}) and (\ref{bound2}), are manifestly a scalar density and a vector density, respectively. (\ref{EHact3}) is the covariant generalization of (\ref{EHact1}).

The Lagrangian density $\mf{L}$ possesses several interesting properties. Taking differentials of the 
identity (\ref{i1}), after a bit of algebra one obtains the relations 
\bea
{\partial \mf{L} \over \partial  \mf{g}^{\mu\nu} }&  = & - \left( \hGamma^\alpha_{\rho\alpha} \hGamma^\rho_{\alpha\nu} - 
\hGamma^\rho_{\mu\nu} \hGamma^\beta_{\rho\beta} \right)    \label{i2}  \\
{\partial \mf{L} \over \partial (\bnabla_\alpha \mf{g}^{\mu\nu}) }  &  = &  -\hGamma^\alpha_{\mu\nu} + {1\over 2} \left( g^\alpha_\nu \hGamma^\rho_{\mu\rho} + g^\alpha_\mu \hGamma^\rho_{\nu\rho} \right)  \, ,  \label{i3}
\eea
from which one may further obtain 
\bea 
\mf{g}^{\mu\nu} {\partial \mf{L} \over \partial  \mf{g}^{\mu\nu} }&  = & -\mf{L}   \label{hom1}\\
(\bnabla_\alpha \mf{g}^{\mu\nu}) {\partial \mf{L} \over \partial (\bnabla_\alpha \mf{g}^{\mu\nu}) } & =  & 2\mf{L} \, . \label{hom2}
\eea
(\ref{hom1}) and (\ref{hom2}) display the homogeneity properties of $\mf{L}$ w.r.t. $\mf{g}^{\mu\nu}$ and $\bnabla_\alpha \mf{g}^{\mu\nu}$, respectively. Adding (\ref{hom1}) and (\ref{hom2}) shows that 
$\mf{L}$ is a homogeneous function of $\mf{g}^{\mu\nu}$ and $\bnabla_\alpha\mf{g}^{\mu\nu}$ of degree $+1$. 
$\mf{L}$ may be equally well considered either a function of $g_{\mu\nu}$, $\bnabla_\alpha g_{\mu\nu}$, or of 
$g^{\mu\nu}$, $\bnabla_\alpha g^{\mu\nu}$, 
and satisfies homogeneity properties with respect to each set. 
In fact, the same homogeneity relations  hold also w.r.t. to the set $g_{\mu\nu}$, $\bnabla_\alpha g_{\mu\nu}$. These properties are easily deduced by considering $g_{\mu\nu}$ of degree $+1$, and, correspondingly, $g^{\mu\nu}$ of degree $-1$ and $\mf{g}^{\mu\nu}$ of degree $+1$. With $\bnabla$ replaced by the ordinary derivative  (\ref{i2}) - (\ref{hom2}) are well-known relations useful in deriving many GR results. They are shown here to generalize in terms of the covariant derivative of an arbitrary symmetric connection as introduced above. 

Expressed in terms of the metric by use of (\ref{barcovder2}) $\mf{L}$ becomes 
\bea 
-\mf{L} & = & {1\over 4} \sqrt{-g} \left[ g^{\alpha\beta} g^{\mu\nu} g^{\rho\sigma} \left( \bnabla_\alpha g_{\mu\rho} \bnabla_\beta g_{\nu\sigma} - \bnabla_\alpha g_{\mu\nu} \bnabla_\beta g_{\rho\sigma} \right)\right. 
\nonumber \\
& & \left. \qquad -2 g^{\mu\nu} g^{\alpha\sigma} g^{\beta\rho} \bnabla_\alpha g_{\mu\rho} \bnabla_\beta g_{\nu\sigma} + 2 g^{\alpha\mu}
g^{\beta\nu} g^{\rho\sigma} \bnabla_\alpha g_{\mu\nu} \bnabla_\beta g_{\rho\sigma} \right]   \, , \label{Ldens3}
\eea 
so that dropping the total divergence the action assumes the form 
\bea
S & = & {1\over \kappa^2} \int d^4x \, [ \sqrt{-g} g^{\mu\nu} \bar{R}_{\mu\nu}   - \mf{L} ]   \nonumber \\
& = &     {1\over \kappa^2} \int d^4x \, \left[ \sqrt{-g} g^{\mu\nu} \bar{R}_{\mu\nu}   
+  (\bnabla_\alpha g_{\mu\nu})\, K^{\alpha\mu\nu\beta\kappa\lambda} (g) \,(\bnabla_\beta g_{\kappa\lambda} ) \right] 
\, , \label{EHact4}
\eea 
where, in symmetrized form,  $K^{\alpha\mu\nu\beta\kappa\lambda} (g)$,  is again given by (\ref{K1}).  (\ref{EHact4}) is then the covariant generalization of (\ref{EHact2}).

So far 
$\bGamma$ has been a symmetric but otherwise arbitrary connection. 
One may specialize to the case where $\bGamma$ is itself the Christoffel of some symmetric rank-2 tensor $\bg$: 
\beq 
\bGamma^\rho_{\mu\nu}(\bg) = {1\over 2} \bg^{\rho\sigma}( \partial_\nu \bg_{\mu\sigma} + \partial_\mu \bg_{\nu\sigma} - \partial_\sigma \bg_{\mu\nu})   \, . \label{bGamma1} 
\eeq  
Most often one is interested in expanding about some background which then serves as the reference metric $\bg$. 
Note that setting $\bg = g$, gives $\hGamma=0$, and (\ref{EHact3}) reduces to (\ref{EHact}), with $\bar{R}=R$, as required for consistency.

Expanding $g_{\mu\nu} = \bg_{\mu\nu} + h_{\mu\nu}$  about a background satisfying the classical field equation $\bar{R}_{\mu\nu} = 0$, the action is simply given by the second term on the r.h.s.  of (\ref{EHact4}). The free part is now 
\beq 
S_2 [\bg, h] = {1\over \kappa^2} \int d^4x \, \left[   
(\bnabla_\alpha h_{\mu\nu})\, K^{\alpha\mu\nu\beta\kappa\lambda} (\bg) \,(\bnabla_\beta h_{\kappa\lambda} ) \right] 
\, ,   \label{FPauli2}
\eeq
and all graviton vertices are obtained by differentiation of it:
\beq 
S_n[\bg,h] = {1\over (n-2)!} \big(\partial_s^{(n-2)} S_2[\bg + s h, h]\big)_{| s=0} \, .  \label{vert2}
\eeq 
This can be more neatly expressed in terms of functional differentiations w.r.t. the background $\bg$: 
\beq
S_n[\bg,h] = {1\over (n-2)!}  \int d^4x \,(\bnabla_\alpha h_{\mu\nu})\, \,(\bnabla_\beta h_{\kappa\lambda} ) \, 
 \left[ \int d^4 y \,  h_{\rho\sigma}(y) {\delta \over \delta \bg_{\rho\sigma}(y) }\right]^{n-2} K^{\alpha\mu\nu\beta\kappa\lambda} (\bg)  \,, \quad (n\geq 3) \,. \label{vert3} 
\eeq
It should be noted that, with $\bg$ and $h$ taken to transform as tensors, each term $S_n$ in this expansion is of manifestly covariant form.  Also, the relation $\Gamma(g)= \bGamma + \hGamma$ is now easily explicitly verified.

Now, if $\bg$ is taken to be the Minkowski metric $\eta$, then $\bGamma=0$, thus $\hGamma= \Gamma(g)$, and the covariant formulation (\ref{EHact4}) reduces to the apparently non-covariant form (\ref{EHact2}) - (\ref{K1}).  The latter can then always be properly viewed as the covariant formulation 
(\ref{EHact4})   
in the special case of a flat connection $\bGamma$, i.e., that for a metric $\bg=\gamma$ satisfying ${\bf Riem}(\gamma)=0$. In general coordinates $\bGamma(\gamma) \not= 0$ and manifest covariance is regained with the covariantized flat space FP action (\ref{FPauli2}) ( i.e., with  $\bg=\gamma$)
as the free kinetic term.

(\ref{EHact4}) gives the most compact form of the GR gravitational action in manifestly covariant form. Correspondingly, 
(\ref{FPauli2}) - (\ref{vert3}) give its most compact form as the theory of a self-interacting massless spin-2 particle. 
To appreciate this better consider the straightforward expansion of the original form of the Einstein-Hilbert action (\ref{EHact}) in powers of the graviton field $h$:  
\beq
S_{\rm\ssc  EH}[\bg+h] = S_{\rm \ssc EH}[\bg] + \sum_{n=2}^\infty S_n[\bg,h]    \label{actexp1} 
\eeq
with 
\beq 
S_n[\bg,h] = {1\over n!} \big(\partial_s^n S_{\rm\ssc  EH}[\bg + s h]\big)_{| s=0} \, .  \label{actexp2}
\eeq 
The linear term $S_1$ is absent when the background satisfies the classical equation of motion 
$\delta S_{\rm EH}[\bg]/\delta \bg_{\mu\nu} =0$. 
We now note that, to within possible total derivative terms, one has 
\beq 
\partial_s S_{\rm \ssc EH}[\bg + s h] = \int d^4x \,h_{\alpha\beta}(x) {\delta \over \delta \bg_{\alpha\beta}(x)} S_{\rm \ssc EH}[\bg + s h] \; . 
\label{actexp3}  
\eeq
By repeated application of (\ref{actexp3}) then, the entire expansion (\ref{actexp1}) can be generated \cite{BdeW1}, \cite{BdeW2}, cf. also \cite{Betal}, from the $S_2[\bg,h]$ quadratic action,  
i.e., one has: 
\beq
S_n[\bg,h] = {2\over n!}\left[ \int d^4x \,h_{\alpha\beta}(x) {\delta \over \delta \bg_{\alpha\beta}(x)}\right]^{n-2}  \, S_2[\bg, h]   \, , \qquad n\geq 3 \,  \label{pactrec} 
\eeq
with
\beq 
S_2[\bg,h] = {1\over 2!}\left[ \int d^4x \,h_{\alpha\beta}(x) {\delta \over \delta \bg_{\alpha\beta}(x)}\right]^2  \, S_{\rm\ssc EH}[\bg]   \, . \label{pactrec0} 
\eeq
Each $S_n[\bg,h]$, $n\geq 2$ generated by (\ref{pactrec}) - (\ref{pactrec0}) \cite{BdeW1} is also of manifestly covariant form with $\bg$ and $h$ transforming as tensors. In carrying out the successive differentiations, however, in addition to terms formed only out of $\bg$, $h$ and $\bnabla h$, 
terms containing factors of $\bar{R}_{\mu\nu}$ and $\bar{R}$ are generated. The background equations of motion cannot be used in them to obtain, say, $S_{n-1}$ until {\it after} the differentiations needed to generate $S_n$ have been carried out. This progressively  results into extremely lengthy expressions. 

This is to be contrasted with the much simpler expressions (\ref{vert3}), where, after the $\bg$ background equation of motion is first immediately used in (\ref{EHact4}), the differentiations involved are those of simple algebraic powers of $\bg$. 
This is a tremendous simplification compared to the result of the direct expansion (\ref{pactrec}).  
It is only after laborious rearrangements of terms, integrations by parts, use of commutation relations of covariant derivatives, etc, that the expressions (\ref{pactrec}) can be shown to be equivalent 
to (\ref{vert3}).\footnote{Instances of this equivalence have on occasion been referred to in some literature as `miraculous', `hidden simplicity' and the like, but, in fact, originate in the ninety year old identification of $\mf{L}$ as the ``equivalent" GR  Lagrangian. \label{Fequiv}} 

We have only considered (\ref{EHact}) in the above discussion, but it is, of course, straightforward to include a cosmological term and/or matter couplings to it. Thus, with a cosmological term,  (\ref{EHact4}) is modified to 
\beq
S =  {1\over \kappa^2} \int d^4x \, \left[ \sqrt{-g} (g^{\mu\nu} \bar{R}_{\mu\nu} - 2\lambda)    
+  (\bnabla_\alpha g_{\mu\nu})\, K^{\alpha\mu\nu\beta\kappa\lambda} (g) \,(\bnabla_\beta g_{\kappa\lambda} ) \right] 
\, . \label{EHact5}
\eeq 
Expansion about a background $\bg$ satisfying $\bar{R}_{\mu\nu} = \lambda \bg_{\mu\nu}$ now gives a surviving contribution from the first term on the r.h.s. This, apart from the classical action contribution $\int \sqrt{-\bg} 2\lambda$, gives the vertices due to the cosmological term interaction for the graviton now propagating on top of this background.

\section{The self-interacting graviton}
\setcounter{equation}{0}
\setcounter{Roman}{0}  
In this section we take the reference metric $\bg = \gamma$ to be a flat background: ${\bf Riem}(\gamma)=0$, and define $\mf{h}^{\mu\nu}= \mf{g}^{\mu\nu} - \sqrt{-\gamma}\,\gamma^{\mu\nu}$.  
The canonical energy-momentum tensor density for the action (\ref{EHact4}) is given by 
\beq 
\kappa^2 \mf{t}^\mu_{\nu} = \delta^\mu_\nu \mf{L} -{\partial \mf{L}\over\partial  (\bnabla_\mu \mf{g}^{\alpha\beta})} \bnabla_\nu\mf{g}^{\alpha\beta}  \,   \label{e-mtens1}
\eeq
with $\mf{g}^{\mu\nu}, \bnabla_\alpha\mf{g}^{\mu\nu}$ chosen as the variable set. (By the homogeneity properties of $\mf{L}$ above, the same expression holds, as can be easily checked,  with this set replaced by $g_{\mu\nu}, \bnabla_\alpha g_{\mu\nu}$.) 
(\ref{e-mtens1}) is, by the equations of motion, conserved: $\bnabla_\mu\mf{t}^\mu_{\nu} = 0$. 

The field equations are in fact equivalent to the following relation obeyed by the canonical energy-momentum tensor: 
\beq 
{ \kappa^2\over 2} \mf{t}^\mu_{\nu} = - \bnabla_\lambda \mf{r}^{\lambda\mu}_{\ \nu}  \, ,  \label{tolman1}
\eeq
where 
\beq
\mf{r}^{\lambda\mu}_{\ \ \nu} \equiv {\partial \mf{L} \over \partial (\bnabla_\lambda\mf{g}^{\nu\alpha} )} \mf{g}^{\mu\alpha} 
- {1\over 2} \delta^\mu_\nu {\partial \mf{L} \over \partial (\bnabla_\lambda\mf{g}^{\alpha\beta} )} \mf{g}^{\alpha\beta} 
\, . \label{tolman2}
\eeq
The non-covariant form of this notable  
relation was first obtained by Tolman \cite{Tol} by explicit computation using the equations of motion. It is known, however, that it can be more simply derived by using the invariance of the action (\ref{EHact4}) under general linear transformations of the coordinates and enforcing the field equations.  
By explicit computation from the definition (\ref{tolman2}) one finds 
\beq
\mf{r}^{\lambda\sigma}_{\ \ \rho} = {1\over 2} \mf{g}_{\kappa\rho} \left( \mf{g}^{\sigma\alpha} \bnabla_\alpha \mf{g}^{\kappa\lambda} - \mf{g}^{\lambda\alpha} \bnabla_\alpha \mf{g}^{\kappa\sigma}  \right) + {1\over 2} \bnabla_\rho \mf{g}^{\sigma\lambda} - 
{1\over 2} \delta^\sigma_\rho \bnabla_\beta \mf{g}^{\beta\lambda}       \, . \label{tolman2a}
\eeq
Note that using (\ref{tolman2a}) in (\ref{tolman1}) immediately implies $\bnabla_\mu\mf{t}^\mu_{\nu} = 0$.

As it is well known, however, for any field system carrying intrinsic angular momentum (nonzero spin) the canonical definition does not yield a symmetric energy-momentum tensor. The standard procedure rendering it symmetric \cite{Bel}, \cite{Rsf}, known as Belinfante symmetrization, is to add to it the divergence of a ``superpotential" which is indeed constructed out of the field spin density. 
The spin tensor density is here given by 
\beq
{\kappa^2\over 2}   \mf{s}^{\mu\nu\lambda} = {\partial \mf{L} \over \partial (\bnabla_\lambda\mf{g}^{\alpha\beta} )} \left( \mf{g}^{\mu\alpha} \gamma^{\nu\beta} - \mf{g}^{\nu\alpha} \gamma^{\mu\beta}  \right)  \, .  \label{spindens1}
\eeq
The appropriate superpotential constructed from the spin density is then 
 \beq
\ms{S}^{\mu\nu\lambda} = -{1\over 2} \left( \mf{s}^{\mu\nu\lambda}  + \mf{s}^{\lambda \mu\nu} + \mf{s}^{\lambda\nu\mu} \right) \, .   \label{Bsuperpot1} 
 \eeq
By explicit computation 
one finds that 
\beq 
{\kappa^2\over 2}\ms{S}^{\mu\nu\lambda} = -{1\over 2}  \gamma^{\mu\rho} \left[ (g_{\rho\sigma}g^{\lambda\beta} - \gamma_{\rho\sigma}\gamma^{\lambda\beta})\bnabla_\beta \mf{g}^{\nu\sigma} - 
(g_{\rho\sigma}g^{\nu\beta} - \gamma_{\rho\sigma}\gamma^{\nu\beta})
\bnabla_\beta \mf{g}^{\lambda\sigma} \right]\, .  \label{Bsuperpot2}
\eeq
Note the antisymmetry property $\ms{S}^{\mu\nu\lambda} = -\ms{S}^{\mu\lambda\nu}$. 
(\ref{Bsuperpot2}) shows that $\ms{S}^{\mu\nu\lambda}$ contains only quadratic and higher powers of the graviton field $\mf{h}^{\mu\nu}$. 

Define now a new energy-momentum tensor density obtained from the canonical one 
by the addition of a $\bnabla \ms{S}$ term: 
\beq 
{\cal T}^\nu_{\,\mu} = \mf{t}^\nu_{\mu} + \bnabla_\lambda \ms{S}^{\nu\lambda}_\mu   \, , 
\label{e-mtens2}
\eeq
where $\ms{S}^{\nu\lambda}_\mu = \gamma_{\mu\sigma} \ms{S}^{\sigma\nu\lambda}$. It  is automatically conserved, $\bnabla_\nu {\cal T}^\nu_{\,\mu} = 0$, by the antisymmetry property of $\ms{S}^{\nu\lambda}_\mu$. Like the canonical $\mf{t}^\nu_\mu$, it contains only quadratic and higher powers in $\mf{h}$.   
We now note that the spin density (\ref{spindens1}) can be expressed in terms of the quantity $\mf{r}^{\lambda\mu}_{\ \ \nu}$ occurring in (\ref{tolman1}): 
\beq
{\kappa^2 \over 2} \mf{s}^{\mu\nu\lambda} =  \mf{r}^{\lambda\mu}_{\ \ \beta} \gamma^{\nu\beta} -   \mf{r}^{\lambda\nu}_{\ \ \beta} \gamma^{\mu\beta}  \, .  \label{spindens2}
\eeq  
Inserting (\ref{spindens2}) in the definition of the superpotential (\ref{Bsuperpot1}) gives
\beq 
{\kappa^2 \over 2} \ms{S}^{\mu\nu\rho} = -{1\over 2} \Bigg[ \gamma^{\nu\alpha} ( \mf{r}^{\rho\mu}_{\ \ \alpha} +  \mf{r}^{\mu\rho}_{\ \ \alpha} ) -  \gamma^{\rho\alpha} ( \mf{r}^{\nu\mu}_{\ \ \alpha} +  \mf{r}^{\mu\nu}_{\ \ \alpha} ) 
+  \gamma^{\mu\alpha} ( \mf{r}^{\nu\rho}_{\ \ \alpha} -  \mf{r}^{\rho\nu}_{\ \ \alpha} )  \Bigg]  \, . \label{Bsuperpot3}
\eeq
Combining (\ref{tolman1}) and (\ref{Bsuperpot3}) with (\ref{e-mtens2}) gives then the Tolman relation in terms of 
the new energy-momentum tensor ${\cal T}^{\mu\nu} = \gamma^{\mu\sigma} {\cal T}^\nu_{\, \sigma}$: 
\beq
{\kappa^2\over 2}  {\cal T}^{\mu\nu} = - {1\over 2} \bnabla_\alpha \Bigg[  \gamma^{\nu\lambda} ( \mf{r}^{\alpha\mu}_{\ \ \lambda} +  \mf{r}^{\mu\alpha}_{\ \ \lambda} )  +  \gamma^{\mu\lambda} ( \mf{r}^{\alpha\nu}_{\ \ \lambda} +  \mf{r}^{\nu\alpha}_{\ \ \lambda} )  - \gamma^{\alpha\lambda} ( \mf{r}^{\nu\mu}_{\ \ \lambda} +  \mf{r}^{\mu\nu}_{\ \ \lambda} ) 
\Bigg]  \, . \label{e-mtens3}
\eeq
As it is manifest from (\ref{e-mtens3}) ${\cal T}^{\mu\nu}$ is a symmetric tensor.

Now, from (\ref{tolman2a}) one gets the explicit simple result: 
\beq
\mf{r}^{\nu\mu}_{\ \ \lambda} + \mf{r}^{\mu\nu}_{\ \ \lambda}= \bnabla_\beta \left( \delta^\beta_\lambda \mf{g}^{\nu\mu} - {1\over 2} \delta^\nu_\lambda \mf{g}^{\mu\beta} - {1\over 2} \delta^\mu_\lambda \mf{g}^{\nu\beta}   \right) \, . \label{tolman3}
\eeq
Combining (\ref{tolman3}) with (\ref{e-mtens3}) finally yields
\beq
{1\over 2}\bnabla_\alpha\bnabla_\beta \left( \gamma^{\alpha\beta}\mf{g}^{\mu\nu} - \gamma^{\nu\beta} \mf{g}^{\mu\alpha} 
- \gamma^{\mu\beta}\mf{g}^{\nu\alpha} + \gamma^{\mu\nu}\mf{g}^{\alpha\beta} \right)  = 
8\pi G  {\cal T}^{\mu\nu} \, . \label{greom1}
\eeq
In (\ref{greom1}) the l.h.s. is linear in the graviton field $\mf{h}^{\mu\nu}$; 
whereas the r.h.s., as already noted, 
contains second and higher powers of $\mf{h}^{\mu\nu}$. Thus, after the customary rescaling $\mf{h} \to \kappa \mf{h}$, giving the graviton field conventional  boson field mass-dimension, the r.h.s has an expansion with the $n$-th order term ($n\geq 2$) 
containing $n$ factors of $\mf{h}$, two $\bnabla$ derivatives and an overall $\kappa^{(n-1)}$ factor.


(\ref{greom1}) recasts the Einstein field equations as the field equation of the spin-2 graviton, the l.h.s. being the free kinetic part and the r.h.s. being the interaction current, which for gravity is, of course, the energy-momentum tensor. The point to be made here is that in (\ref{greom1}) this tensor is given precisely by the standard field-theoretic prescription, i.e., the canonical energy-momentum tensor with the necessary Belinfante-symmetrization appropriate for a spin-2 particle. 
(\ref{greom1}) was first obtained, in non-covariant form, by Papapetrou \cite{Pp} long ago, from a different perspective than the present-day, quantum-motivated view of the propagating graviton on a background. 


It is straightforward to include matter in the above. All that is needed is to replace $\mf{t}^\mu_\nu$ by $\mf{t}^\mu_\nu + \mf{T}^\mu_\nu$ in (\ref{tolman1}) above, which then gives the general Tolman relation including matter. Here 
$\mf{T}^\mu_\nu = \sqrt{-g} T^\mu_{{\rm M}\, \nu}$ is the matter energy-momentum tensor density. Correspondingly, one defines 
\beq 
{\cal T}^\nu_{\,\mu} = \mf{t}^\nu_{\mu} + \mf{T}^\mu_\nu +  \bnabla_\lambda \ms{S}^{\nu\lambda}_\mu   \,  
\label{e-mtens4}
\eeq
and it is (\ref{e-mtens4}) that is now used in (\ref{e-mtens3}) and (\ref{greom1}).

(\ref{greom1}) appears to be the most concise version of the graviton field equation achieved by use of $\mf{g}^{\mu\nu}$ as the basic variables. 
This is not, however, how it is usually presented in the literature. 
The customary presentation, e.g., \cite{W}, simply rewrites the Einstein field equation by keeping the linearized part of $(R^{\mu\nu} - {1\over 2} g^{\mu\nu}R)$ on the l.h.s. and shuffling the rest onto the r.h.s., which then defines the graviton field pseudo-tensor. This has the advantage of being automatically symmetric. It seems, however, to bear no relation to the canonical field energy-momentum (pseudo)tensor; in fact, the suggestion seems to be that no such connection exists. Having established (\ref{greom1}), however, it should be that such different representations
reflect the inherent ambiguity in any local definition of field energy-momentum density, and thus be equivalent to within the addition of superpotentials. 

To make this connection we revert from $\mf{g}^{\mu\nu}$ to $g_{\nu\mu}$ as our basic variables.  In terms of the latter one finds that, after some algebra, (\ref{greom1}) can be re-expressed  as follows. Let 
\bal
\tilde{\cal T}^{\mu\nu}  &= \gamma^{\mu\lambda} (\mf{t}^\nu_{\lambda} + \mf{T}^\mu_\lambda)  + \bnabla_\lambda \ms{S}^{\mu\nu\lambda} + \bnabla_\lambda \ms{S}_{\rm I}^{\mu\nu\lambda}  \nonumber \\
& = {\cal T}^{\mu\nu} + \bnabla_\lambda \ms{S}_{\rm I}^{\mu\nu\lambda}   \label{e-mtens5}
\end{align}  
be a redefined energy-momentum tensor density that differs from (\ref{e-mtens4}) by the addition of the superpotential
\bal
{\kappa^2\over 2} \ms{S}_{\rm I}^{\mu\nu\alpha} = {1\over 2} \sqrt{-\gamma} &   \Bigg[   \left({\sqrt{-g}\over \sqrt{-\gamma}}\right)  g^{\mu\rho} ( \gamma^{\alpha\beta}g^{\nu\sigma} -\gamma^{\nu\beta} g^{\alpha\sigma}) - \gamma^{\mu\rho} ( \gamma^{\alpha\beta}\gamma^{\nu\sigma} -\gamma^{\nu\beta} \gamma^{\alpha\sigma})  \nonumber \\
&   +  \left({\sqrt{-g}\over \sqrt{-\gamma}}\right)  g^{\beta\sigma} ( \gamma^{\mu\nu}g^{\alpha\rho} -\gamma^{\mu\alpha} g^{\nu\rho}) - \gamma^{\beta\sigma} ( \gamma^{\mu\nu}\gamma^{\alpha\rho} -\gamma^{\mu\alpha} \gamma^{\nu\rho})  \nonumber  \\
&  
-{1\over 2} \left({\sqrt{-g}\over \sqrt{-\gamma}}\right)  g^{\rho\sigma} (\gamma^{\alpha\beta} g^{\mu\nu} -\gamma^{\nu\beta} g^{\mu\alpha} + \gamma^{\mu\nu}g^{\alpha\beta} - \gamma^{\mu\alpha}g^{\nu\beta} )   \nonumber \\
&   +    \gamma^{\rho\sigma} (\gamma^{\alpha\beta} \gamma^{\mu\nu} -\gamma^{\nu\beta} \gamma^{\mu\alpha} )  \Bigg]  
 \bnabla_\beta g_{\rho\sigma}  \, .   \label{superpotII}
\end{align} 
$\tilde{\cal T}^{\mu\nu}$ is symmetric and conserved. 
(\ref{greom1}) can then be written in the form 
\beq
2  K^{\alpha\mu\nu\beta\rho\sigma}(\gamma) \bnabla_\alpha \bnabla_\beta\, g_{\rho\sigma} =
- 8\pi G \tilde{\cal T}^{\mu\nu}    \, .     \label{greom2}
\eeq
The l.h.s. of (\ref{greom2}) is now the covariantized Fierz-Pauli  free kinetic part, 
the  appropriate self-interactions for which are given on the r.h.s. by the energy-momentum tensor density (\ref{e-mtens5}). 
Setting $\gamma=\eta$ in (\ref{greom2}) gives precisely what one gets by putting the linearized part of $(R^{\mu\nu} - {1\over 2} g^{\mu\nu}R)$ on the l.h.s. and all the rest onto the r.h.s. It is thus seen that this procedure amounts to modifying the symmetrized canonical energy-momentum tensor density by the addition of an appropriate superpotential term.

The expression for $\mf{r}^{\mu\sigma}_{\ \ \rho}$ given by (\ref{tolman2a}) may be rewritten in the form 
\beq
\mf{r}^{\mu\sigma}_{\ \ \rho} =  {1\over 2} \mf{g}_{\nu\rho} \bnabla_\alpha U^{\mu\sigma\nu\alpha} + {1\over 2} \bnabla_\alpha
\left(\delta^\alpha_\rho \mf{g}^{\sigma\mu} - \delta^\mu_\rho \mf{g}^{\sigma\alpha}\right)  \, , \label{tolman4}
\eeq
where 
\beq
U^{\mu\sigma\nu\alpha} = \left( \mf{g}^{\mu\nu} \mf{g}^{\sigma\alpha} - \mf{g}^{\mu\alpha}\mf{g}^{\sigma\nu} \right) \,.
\label{tolman5}
\eeq
Note that $U^{\mu\sigma\nu\alpha} = U^{[\mu\sigma][\nu\alpha]}$. Using (\ref{tolman4}) in (\ref{tolman1}) gives the Tolman equation (including matter and after raising the covariant index by the metric density $\mf{g}$) in the alternative form 
\beq 
- \kappa^2 \mf{g}^{\rho\lambda}(\mf{t}^\sigma_\lambda + \mf{T}^\sigma_\lambda) =\mf{g}^{\rho\lambda}  \bnabla_\mu \left( \mf{g}_{\nu\lambda} \bnabla_\alpha U^{\mu\sigma\nu\alpha} \right)  \, .  \label{tolman6}
\eeq
Thus, if we now define 
\beq 
{\cal T}^{\rho\sigma}_{\ssc{\rm LL}} = \mf{g}^{\rho\lambda} \mf{t}^\sigma_\lambda  - {1\over \kappa^2} (\bnabla_\mu \mf{g}^{\rho\lambda})  \left( \mf{g}_{\nu\lambda} \bnabla_\alpha U^{\mu\sigma\nu\alpha} \right)  \, , \label{LL1}
\eeq 
(\ref{tolman6}) gives 
\beq 
{\cal T}^{\rho\sigma}_{\ssc{\rm LL}}  + \mf{g}^{\rho\lambda} \mf{T}^\sigma_\lambda = {1\over \kappa^2} \bnabla_\mu\bnabla_\alpha U^{\sigma\mu\rho\alpha}  \, .  \label{LL2}
\eeq
${\cal T}^{\rho\sigma}_{\ssc{\rm LL}}$ is the LL gravitational energy-momentum density, and 
(\ref{LL2}) is the LL form of the field equations given in terms of ${\cal T}^{\rho\sigma}_{\ssc{\rm LL}}$ \cite{LL}.  
With symmetric matter tensor $\mf{g}^{\rho\lambda} \mf{T}^\sigma_\lambda = (-g) T^{\rho\sigma}_M$, it is  seen from 
(\ref{LL2}) that ${\cal T}^{\rho\sigma}_{\ssc{\rm LL}}$ is indeed symmetric and that 
$\bnabla_\rho ({\cal T}^{\rho\sigma}_{\ssc{\rm LL}}  + \mf{g}^{\rho\lambda} \mf{T}^\sigma_\lambda)=0$. As seen from (\ref{LL1}) and (\ref{e-mtens1}) ${\cal T}^{\rho\sigma}_{\ssc{\rm LL}}$ contains only first derivatives of the metric and can be explicitly computed from these relations.

(\ref{LL2}) was originally derived \cite{LL} in a rather different manner that makes no reference to the canonical energy-momentum tensor density. Here we arrived at it via the Tolman equation (\ref{tolman6}) and obtained an explicit relation, (\ref{LL1}), between ${\cal T}^{\rho\sigma}_{\ssc{\rm LL}}$ and 
$\mf{g}^{\rho\lambda} \mf{t}^\sigma_\lambda$, 
i.e the canonical density with the covariant index raised by the metric density $\mf{g}$. This raising by $\mf{g}$ in the LL definitions \cite{LL}, it is 
important to note, leads to ${\cal T}^{\rho\sigma}_{\ssc{\rm LL}}$  being a tensor density of weight 2.  
One may define a corresponding tensor by letting ${\cal T}^{\rho\sigma}_{\ssc{\rm LL}} \equiv (-g) t^{\rho\sigma}_{\ssc{\rm LL}}$, but it is the weight 2 densities, ${\cal T}^{\rho\sigma}_{\ssc{\rm LL}}$ for pure gravity and  $(-g) (t^{\rho\sigma}_{\ssc{\rm LL}}  + T^{\rho\sigma}_M)$ in the presence of matter, that enter (\ref{LL2}) and are conserved.

This is actually something of a problem if, in asymptotically flat spacetimes, one defines the total field momentum as the spatial integral of $(-g) (t^{\rho 0}_{\ssc{\rm LL}}  + T^{\rho 0}_M)$, which can then be evaluated over a 2-sphere at spatial infinity, since it would not have the expected vector transformation property. 
To correct this when working with the LL tensor and revert to tensor densities of weight 1  one needs a reference metric, such as $\gamma$, to form ${\cal T}^{\rho\sigma}_{\ssc{\rm LL}}/\sqrt{-\gamma}$.\footnote{This can always be done in formulations employing a reference metric, as, e.g., was done in \cite{PP}. } 
In the formulation presented here this is naturally arrived at  since we have the relation 
\beq
{ 1 \over 2} \bnabla_\alpha\bnabla_\beta U^{\iota\alpha\kappa\beta} = {1\over 2} \sqrt{-\gamma} \, 
\bnabla_\alpha\bnabla_\beta 
\left[ \mf{g}^{\iota\kappa} \gamma^{\alpha\beta} + \mf{g}^{\alpha\beta}\gamma^{\iota\kappa} - \mf{g}^{\iota\alpha} \gamma^{\kappa\beta} - \mf{g}^{\kappa\alpha} \gamma^{\iota\beta}   \right]   + 
{\kappa^2\over 2} \bnabla_\alpha \ms{S}_{\ssc{\rm LL}}^{\iota\kappa\alpha}  \, , \label{emrel1}
\eeq
where
\beq 
\kappa^2\ms{S}_{\ssc{\rm LL}}^{\iota\kappa\alpha}  =  \bnabla_\beta \left[ \mf{h}^{\iota\kappa} \mf{h}^{\alpha\beta} - \mf{h}^{\iota\alpha} \mf{h}^{\kappa\beta} \right] \, . \label{LL3}
\eeq
Comparing (\ref{emrel1}) with (\ref{greom1}) and (\ref{LL2}) in the pure gravity case one obtains 
\beq 
{1\over \sqrt{-\gamma}} {\cal T}^{\iota\kappa}_{\ssc{\rm LL}}  =  {\cal T}^{\iota\kappa} + \bnabla_\alpha \Big( {1\over \sqrt{-\gamma}} \ms{S}_{\ssc{\rm LL}}^{\iota\kappa\alpha}\Big) \, .  \label{emrel2}
\eeq
(\ref{emrel2}) explicitly relates the (reweighted) LL and the Belinfante-symmetrized canonical energy-momentum tensors by a superpotential term. 

The explicit formula for  ${\cal T}^{\rho\sigma}_{\ssc{\rm LL}}$, known from the original derivation \cite{LL}, may be directly obtained from our relation (\ref{LL1}) and (\ref{e-mtens1}). Similarly, an explicit expression for ${\cal T}^{\iota\kappa}$ may be obtained from (\ref{e-mtens1}) and (\ref{Bsuperpot2}). Alternatively, using (\ref{emrel2}) one finds  
\bal
{\cal T}^{\iota\kappa} &={1\over \kappa^2} {1\over \sqrt{-\gamma}} \Bigg[   (\bnabla_\beta\mf{h}^{\iota\alpha}) (\bnabla_\alpha \mf{h}^{\kappa\beta})  - (\bnabla_\alpha\mf{h}^{\iota\kappa}) (\bnabla_\beta \mf{h}^{\alpha\beta})  
 + {1\over 2}  g^{\iota\kappa}g_{\lambda\rho} (\bnabla_\gamma\mf{h}^{\lambda\sigma})(\bnabla_\sigma\mf{h} ^{\gamma\rho})  \nonumber \\
&- \Big( g^{\iota\lambda} g_{\rho\sigma} (\bnabla_\gamma\mf{h}^{\kappa\sigma})(\bnabla_\lambda\mf{h} ^{\rho\gamma}) 
+ g^{\kappa\lambda} g_{\rho\sigma} (\bnabla_\gamma\mf{h}^{\iota\sigma})(\bnabla_\lambda\mf{h} ^{\rho\gamma}) \Big) 
+ g_{\lambda\rho} g^{\sigma\gamma} (\bnabla_\sigma\mf{h}^{\iota\lambda})(\bnabla_\gamma\mf{h} ^{\kappa\rho}) \nonumber \\
& +{1\over 8} ( 2g^{\iota\lambda}g^{\kappa\rho} - g^{\iota\kappa} g^{\lambda\rho}) 
( 2g_{\sigma\gamma}g_{\delta\epsilon} - g_{\gamma\delta} g_{\sigma\epsilon}) (\bnabla_\lambda\mf{h}^{\sigma\epsilon})(\bnabla_\rho\mf{h} ^{\gamma\delta})   \nonumber \\
& + \mf{h}^{\kappa\beta}\bnabla_\alpha\bnabla_\beta\mf{h}^{\iota\alpha} + 
\mf{h}^{\iota\alpha}\bnabla_\alpha\bnabla_\beta\mf{h}^{\kappa\beta} 
- \mf{h}^{\alpha\beta}\bnabla_\alpha\bnabla_\beta\mf{h}^{\iota\kappa} - 
\mf{h}^{\iota\kappa}\bnabla_\alpha\bnabla_\beta\mf{h}^{\alpha\beta} 
 \Bigg]  \, . \label{e-mtens6}
\end{align}

All our discussion above has been carried in a gauge independent manner. In explicit calculations one frequently has to specify a gauge.  
A common, and here a natural choice is the de Donder gauge
\beq
\bnabla_\nu \mf{g}^{\mu\nu} = 0  \, ,  \label{dDgauge}
\eeq
in which (\ref{greom1}) assumes the noteworthy form
\bal 
\bar{\Box} \,\mf{h}^{\mu \nu} 
& = {1\over \sqrt{-\gamma}} \Bigg[   (\bnabla_\beta\mf{h}^{\mu\alpha}) (\bnabla_\alpha \mf{h}^{\nu\beta})   
 + {1\over 2}  g^{\mu\nu}g_{\lambda\rho} (\bnabla_\gamma\mf{h}^{\lambda\sigma})(\bnabla_\sigma\mf{h} ^{\gamma\rho})  
 - \mf{h}^{\alpha\beta}\bnabla_\alpha\bnabla_\beta\mf{h}^{\mu\nu}
 \nonumber \\
&- \Big( g^{\mu\lambda} g_{\rho\sigma} (\bnabla_\gamma\mf{h}^{\nu\sigma})(\bnabla_\lambda\mf{h} ^{\rho\gamma}) 
+ g^{\nu\lambda} g_{\rho\sigma} (\bnabla_\gamma\mf{h}^{\mu\sigma})(\bnabla_\lambda\mf{h} ^{\rho\gamma}) \Big) 
+ g_{\lambda\rho} g^{\sigma\gamma} (\bnabla_\sigma\mf{h}^{\mu\lambda})(\bnabla_\gamma\mf{h} ^{\nu\rho}) \nonumber \\
& +{1\over 8} ( 2g^{\mu\lambda}g^{\nu\rho} - g^{\mu\nu} g^{\lambda\rho}) 
( 2g_{\sigma\gamma}g_{\delta\epsilon} - g_{\gamma\delta} g_{\sigma\epsilon}) (\bnabla_\lambda\mf{h}^{\sigma\epsilon})(\bnabla_\rho\mf{h} ^{\gamma\delta})   
 \Bigg]  \, . \label{greom3}
\end{align}
(\ref{greom3}) gives perhaps the most compact expression in closed form of the entire dynamical content of GR  as the field equation of the graviton with the interaction current being its symmetrized canonical energy-momentum tensor. It is the exact analog of the Lorentz gauge photon field equation $\Box A^\mu = J^\mu$.

\section{Conclusion}
\setcounter{equation}{0}
\setcounter{Roman}{0} 

Working with the Lagrangian density $\mf{L}$ (the ``$\Gamma\Gamma$" form), rather than the original EH form of the Lagrangian,  has long been known to provide efficient derivations and compact expressions for many GR results. This becomes particularly relevant in perturbative quantum gravity where it provides the most concise way of generating the graviton vertices.  The density $\mf{L}$, however, is not an invariant density. In this note we have seen how it can be covariantized by the introduction of a reference connection. This arbitrary connection may, in particular, be taken to be a metric connection, thus introducing a reference metric. 
Reference (flat) metrics for this purpose have been used before. e.g., \cite{PP}. 
The present scheme based on (\ref{barcovder1})-(\ref{barcovder2}) and (\ref{Riem1}) is straightforward and can be applied to any metric gravity theory. 
It naturally includes, as a special case, the usual setup of propagation over a background. 
For perturbative gravity, a simple compact formula for the generation of graviton vertices which is manifestly background field invariant can then be given. This was contrasted with the formula for generating vertices obtained by expansion of the EH action in a manner preserving background field invariance. The latter generally leads to much lengthier expressions that can be only laboriously reduced to those obtained from the former.

We next showed how the Einstein field equations may be recast as the field equation of the spin-2 graviton coupled to the gravitational canonical energy-momentum tensor. 
The canonical definition of energy-momentum tensor is not symmetric for any field carrying intrinsic angular momentum, i.e.,  non-zero spin. A symmetric version can be constructed by the addition of a superpotential term in a construction that applies to any spin \cite{Bel}, \cite{Rsf}. 
This allows one to  exhibit the graviton self-interactions by following standard field theory procedures. The construction,  
based on older work in \cite{Tol} and \cite{Pp}, was carried out in the present covariantized framework with flat background metric. Other definitions of the gravitational field energy-momentum (pseudo)tensor  were then shown to be related to the canonical version by the addition of appropriate superpotential terms. This was done in particular 
for the standard textbook pseudotensor defined as the nonlinear part of the Einstein tensor, and the LL energy-momentum tensor. We gave the exact relation between the commonly employed LL  tensor and the canonical definition, as well as the LL and the Belinfante-symmetrized canonical tensors, and, in all cases,  the explicit form of the appropriate superpotentials. These have not appeared in the literature before. The use of the metric density $\mf{g}$ variable in obtaining relatively concise formulas in closed form, such as (\ref{e-mtens6}) and (\ref{greom2}), should be noted in this connection.

Showing explicitly how various definitions of the local gravitational energy-momentum are equivalent, 
just as in any other field theory, by the addition of appropriate superpotentials makes it apparent that they all  must lead to the same total field 4-momentum for asymptotically flat spacetimes. The covariant formulation given here  makes it also apparent that this agreement is independent of the coordinates used. These points, the source of some confusion in the literature over the years, are hopefully clarified by the explicit results presented here.

Our exploration of the self-interacting graviton naturally suggests the following question. As demonstrated by the general argument in \cite{De1} consistent self-coupling of a massless spin-2 particle leads to GR. Equation (\ref{greom1}) above, in particular, expresses this in a form complying to the precepts of standard field theory, the interaction current being precisely the symmetrized canonical energy-momentum tensor. One may then ask what modifications may be introduced in the above framework that would lead to a theory of the self-interacting graviton truly beyond GR. The only possibility would seem to be 
to give up the assumption of locality of interactions. There are in fact several physical reasons for considering the delocalization of graviton interactions.  This will be considered elsewhere.






\end{document}